\documentclass[10pt, conference, compsocconf]{IEEEtran}
\usepackage{cite}
\usepackage{amsmath,amssymb,amsfonts}
\usepackage{algorithmic}
\usepackage{graphicx}
\usepackage{textcomp}
\usepackage{xcolor}
\usepackage{booktabs}
\usepackage{colortbl}
\usepackage{flushend}
\usepackage{footnote}
\makesavenoteenv{tabular}
\makesavenoteenv{table}

\def\BibTeX{{\rm B\kern-.05em{\sc i\kern-.025em b}\kern-.08em
    T\kern-.1667em\lower.7ex\hbox{E}\kern-.125emX}}

\RequirePackage[hyphens]{url}
\usepackage{listings}
\usepackage[hidelinks, bookmarks=false]{hyperref}
\usepackage{color}
\definecolor{lightgray}{rgb}{.9,.9,.9}
\definecolor{darkgray}{rgb}{.4,.4,.4}
\definecolor{purple}{rgb}{0.65, 0.12, 0.82}

\lstdefinelanguage{diff}{
  morecomment=[f][\color{blue}]{@@},     % group identifier
  morecomment=[f][\color{red}]-,         % deleted lines
  morecomment=[f][\color{green}]+,       % added lines
  morecomment=[f][\color{magenta}]{---}, % Diff header lines (must appear after +,-)
  morecomment=[f][\color{magenta}]{+++},
}

\lstdefinelanguage{JavaScript}{
  keywords={typeof, new, true, false, catch, function, return, null, catch, switch, var, if, in, while, do, else, case, break},
  keywordstyle=\color{blue}\bfseries,
  ndkeywords={class, export, boolean, throw, implements, import, this},
  ndkeywordstyle=\color{darkgray}\bfseries,
  identifierstyle=\color{black},
  sensitive=false,
  comment=[l]{//},
  morecomment=[s]{/*}{*/},
  commentstyle=\color{purple}\ttfamily,
  stringstyle=\color{red}\ttfamily,
  morestring=[b]',
  morestring=[b]"
}

\let\OLDthebibliography\thebibliography
\renewcommand\thebibliography[1]{
  \OLDthebibliography{#1}
  \setlength{\parskip}{0pt}
  \setlength{\itemsep}{0pt plus 0.3ex}
}

\begin{document}

%\title{Predicting Security Vulnerabilities in JavaScript Functions Based on Static Source Code Metrics}
\title{Challenging Machine Learning Algorithms in Predicting Vulnerable JavaScript Functions\vspace{-20pt}}

\author{
\IEEEauthorblockN{Rudolf Ferenc\IEEEauthorrefmark{2}, P\'eter Heged\H{u}s\IEEEauthorrefmark{1}, P\'eter Gyimesi\IEEEauthorrefmark{2}, G\'abor Antal\IEEEauthorrefmark{2}, D\'enes B\'an\IEEEauthorrefmark{2}, and Tibor Gyim\'othy\IEEEauthorrefmark{1}\IEEEauthorrefmark{2}}
\IEEEauthorblockA{\IEEEauthorrefmark{1}\textit{MTA-SZTE Research Group on Artificial Intelligence},
Szeged, Hungary \\
\{hpeter $|$ gyimothy\}@inf.u-szeged.hu}
\IEEEauthorblockA{\IEEEauthorrefmark{2}\textit{Department of Software Engineering, University of Szeged},
Szeged, Hungary \\
\{ferenc $|$ pgyimesi $|$ antal $|$ zealot\}@inf.u-szeged.hu}
\vspace{-20pt}
}

\maketitle

\begin{abstract}
The rapid rise of cyber-crime activities and the growing number of devices threatened by them place software security issues in the spotlight.
As around 90\% of all attacks exploit known types of security issues, finding vulnerable components and applying existing mitigation techniques is a viable practical approach for fighting against cyber-crime.
In this paper, we investigate how the state-of-the-art machine learning techniques, including a popular deep learning algorithm, perform in predicting functions with possible security vulnerabilities in JavaScript programs.

We applied 8 machine learning algorithms to build prediction models using a new dataset constructed for this research from the vulnerability information in public databases of the Node Security Project and the Snyk platform, and code fixing patches from GitHub.
We used static source code metrics as predictors and an extensive grid-search algorithm to find the best performing models.
We also examined the effect of various re-sampling strategies to handle the imbalanced nature of the dataset.

The best performing algorithm was KNN, which created a model for the prediction of vulnerable functions with an F-measure of 0.76 (0.91 precision and 0.66 recall).
Moreover, deep learning, tree and forest based classifiers, and SVM were competitive with F-measures over 0.70.
Although the F-measures did not vary significantly with the re-sampling strategies, the distribution of precision and recall did change.
No re-sampling seemed to produce models preferring high precision, while re-sampling strategies balanced the IR measures.

\end{abstract}


%JavaScript has a real traction as a full-stack development language as well as an IoT programming solution, but it is quite neglected in terms of vulnerability or even bug prediction studies.
%Using the vulnerability information available in public databases of the Node Security Project and the Snyk platform and code fixing patches from GitHub, we create a vulnerability dataset of 12, 125 JavaScript functions from which 1496 was vulnerable.

%Given the highly dynamic nature of JavaScript, we got more than encouraging results using only static metric predictors.
\vspace{-5pt}
\section{Introduction}\label{sec:introduction}

%JavaScript is getting traction not just in client-side web development but as a desktop and server language (Node.js) or even as an IoT (e.g. JerryScript~\cite{jerry} or the Espruino framework~\cite{espruino}) implementation language.
JavaScript is getting traction not just in client-side web development but as a desktop and server language (Node.js), mobile app language (React Native), or even as an IoT (e.g. JerryScript or the Espruino framework) implementation language.
Therefore, programs written in JavaScript are exposed more and more to various security risks.

Even though the rapid rise of cyber-crime activities and the growing number of devices threatened by them place software security issues in the spotlight, security concerns of programs are still neglected from time to time.
According to past studies~\cite{mead2010software}, around 90\% of all attacks exploit known types of security issues.
Therefore, finding vulnerable components for applying existing mitigation techniques on them might be a viable practical approach for fighting against cyber-crime.
In this paper, we investigate how the state-of-the-art machine learning techniques~--~including a popular deep learning algorithm~--~perform in predicting functions with possible security vulnerabilities in JavaScript programs.

Security vulnerabilities are very similar to bugs (i.e. most of them can be seen as special types of bugs, though not functional), however, many studies show that bug prediction models cannot be applied for finding vulnerabilities as is~\cite{Shin2011CanTF, zimmermann2010searching}.
Although this suggests that specific prediction models are needed for finding vulnerable software components, we can still leverage the abundance of knowledge already accumulated in the area of bug prediction.
JavaScript, however, is not well studied in terms of bug prediction, so general conclusions based on other languages might not hold.

Moreover, most of the bug prediction models find fault-prone files or classes~\cite{jimenez2018enabling, neuhaus2007predicting,shin2011evaluating,yu2018harmless,
Morrison2015ChallengesWA,chowdhury2011using, zimmermann2010searching}, while rarely working at a finer granularity level (e.g. for methods, functions, or statements~\cite{shin2008empirical}).
%\todoi{ide keresni}
These approaches are less effective for JavaScript, as source code is often structured in only several files (even into one single \texttt{js} file) and usually there are no higher level logical constructs (like classes) above functions.
Prediction models for vulnerable source files would not be really useful in such contexts; we need at least function level vulnerability information and prediction models.

To the best of our knowledge, there are no existing vulnerability datasets specifically for JavaScript programs, which would contain vulnerability information at the level of functions.
VulinOSS~\cite{gkortzis2018vulinoss} and VulData7~\cite{jimenez2018enabling} are very useful proposals with the aim of collecting general vulnerability datasets together with fixing patches.
However, they are not specific to JavaScript and do not map the fixed vulnerabilities to individual functions.

For this study, we created a fine-grained, public, JavaScript vulnerability dataset with data extracted from \emph{nsp} (Node Security Platform~\cite{nspweb}) and the \emph{Snyk Vulnerability Database}~\cite{snyk_vuldb} automatically matched with information available on GitHub (i.e fixing commits and patches).
The new function level vulnerability dataset contains 12,125 functions from which 1,496 are vulnerable.
It includes static code metrics provided by the OpenStaticAnalyzer~\cite{osaweb} and escomplex~\cite{escomplexweb} tools, too.

With the help of this dataset, we investigate if predicting vulnerable functions is feasible based on the fast and easily calculable static code metrics.
We compare the performances of the most widely used machine learning algorithms on this prediction task, including two deep neural network variants, the K-Nearest Neighbors algorithm (KNN), a decision tree classifier (Tree), the C-Support Vector Classification variant of the Support Vector Machine algorithm (SVM), Random Forest (Forest), Logistic regression (Logistic), Linear regression (Linear) and the Gaussian Naive Bayes algorithm (Bayes).
We apply various re-sampling strategies to handle the imbalanced nature of the dataset.

In this paper, we address the following research questions:% in this paper:

\textbf{RQ1:} Is predicting vulnerable JavaScript functions feasible using static source code metrics?

\textbf{RQ2:} How do the various machine learning algorithms perform compared to each other for vulnerability prediction?

%\textbf{RQ3:} How do the different parameters and sampling methods affect the performances of the various machine learning algorithms on vulnerability prediction?

Given the highly dynamic nature of JavaScript, we got encouraging results using only static code metrics as predictors.
%The best performing algorithm was KNN, which created a model for the prediction of vulnerable functions with an F-measure of 0.76 (0.91 precision and 0.66 recall).
%Moreover, deep learning, tree and forest based classifiers, and SVM were competitive with F-measures over 0.70.
%Although F-measures did not vary significantly with the re-sampling strategies, the distribution of precision and recall did change.
%No re-sampling seemed to produce models preferring high precision, while re-sampling strategies balanced the IR measures.
The main contributions of the paper are two-fold:

\begin{itemize}
	\item We release a new public vulnerability dataset consisting of the static analysis results of 12,125 JavaScript functions complemented with the information whether the functions contain a vulnerability or not.
	\item We publish a comprehensive comparison of 8 well-known machine learning algorithms on predicting vulnerable functions.%The best hyper parameters of the algorithms are found by a grid-search algorithm.
	%\item We explicitly analyze the effect of various random under and over-sampling strategies on the prediction capabilities of the machine learning algorithms.
\end{itemize}

%The rest of the paper is organized as follows.
%In Section~\ref{sec:related}, we summarize the works that are related to this paper.
%We describe the vulnerability dataset and data analysis approach in Section~\ref{sec:methodology}.
%The data analysis results are then presented in Section~\ref{sec:results}.
%We overview the possible threats of our work in Section~\ref{sec:threats}, and conclude the paper in Section~\ref{sec:conclusions}.

\begin{figure*}
\centering
\includegraphics[width=1.25\columnwidth]{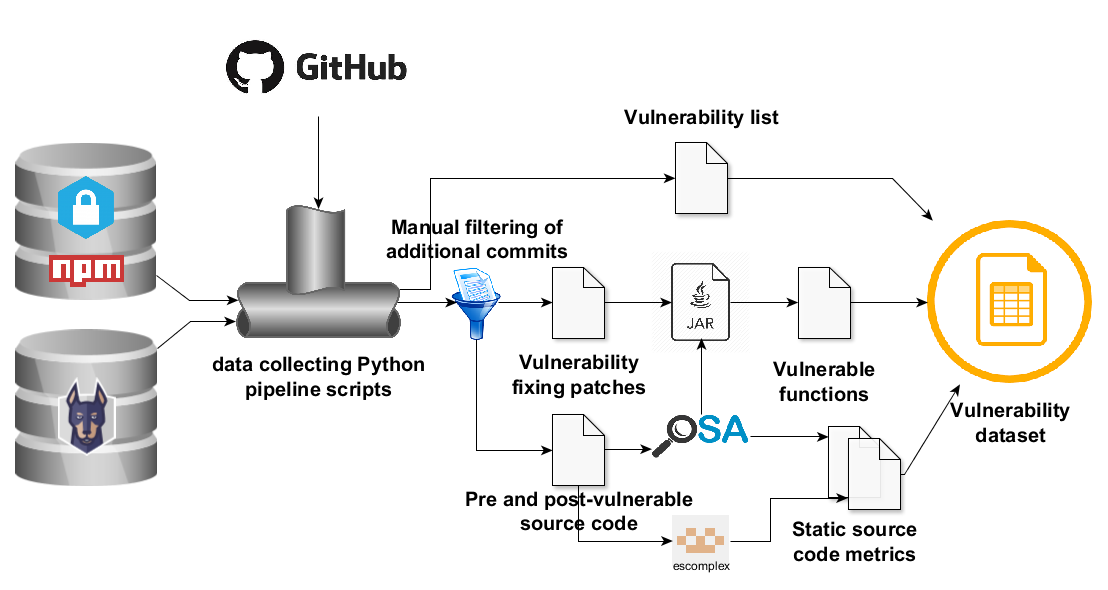}
\vspace{-8pt}
\caption{Data processing overview}
\label{fig:process_overview}
\vspace{-15pt}
\end{figure*}

\vspace{-6pt}
\section{Related Work}\label{sec:related}

% VulData7, VulinOSS
% A vul prediction datasetek altalanosak, nem JS specifikusak es nincs forraskod elemhez mapolva a patch
% Function szintu bug/vuln prediction van-e? Keves ezert ez itt ertekes (plane JS eseteben, ahol a fajl szintu predikcio gaz, osztalyok nem nagyon vannak, stb.)
% Bug vs vuln prediction osszehasonlito cikkek, nem lehet a bug predictiont csak ugy siman alkalmazni vul pred-re, specifikus modellek kellenek
% Bug prediction tul tag, hogy az egeszet atnezzuk, ezert a function levelre ill. a vuln-al valo osszehasonlitasra szoritkoztunk

%\emph{Vulnerability prediction from static source code metrics}.
%Malhotra~\cite{MALHOTRA2015504} concluded in his systematic literature review that machine learning techniques have the ability to predict software faults.

In their preliminary study, Siavvas et al. \cite{siavvaspreliminary} investigated if a relationship exists among software metrics and specific vulnerability types.
They used 13 metrics and found that software metrics may not be sufficient indicators of specific vulnerability types, but using novel metrics could help.
In our study, we used 35 static source code metrics, including various Halstead variants, and found that they can effectively predict vulnerable functions in JavaScript.

In their work, Jimenez et al.~\cite{jimenez2018enabling} proposed an extensible framework (VulData7) and dataset of real vulnerabilities, automatically collected from software archives.
Although it is similar to our work, VulData7 is general, i.e., it contains vulnerabilities for various languages at file level.
%They presented the capabilities of their framework on 4 large systems, but one can extend the framework to meet specific needs.
%Their proposed framework does not contain any source code metrics and there is no description on how to integrate any particular tool into it which could provide information like static source code metrics.
%VulData7 is a general purpose framework, but we could not adapt it as it is not fine-grained since all data about a vulnerability is available only at file level.
Even though it contains JavaScript vulnerabilities, using them in our study was infeasible, as JavaScript files could contain lots of functions.
Our proposed database is more fine-grained, every piece of information is available at the function level, thus enabling more accurate experiments.

Neuhas~et~al.~\cite{neuhaus2007predicting} introduced a new approach (and the corresponding tool) called Vulture, which can predict vulnerable components in the source code, mainly relying on the dependencies between the files.
They analyzed the Mozilla code base to evaluate their approach using SVM for classification.
%A component in their approach is a header-source pair (where both are available, source file otherwise).
%They analyzed imports and function calls between components and used SVM for classification.
Although their results are very promising, we could not locate the proposed tool online.
Contrary to this work, we predict vulnerabilities at the level of JavaScript functions and apply multiple machine learning approaches. %not just SVM.

In their work, Shin~et~al.~\cite{shin2008empirical} created an empirical model to predict vulnerabilities from source code complexity metrics.
Their model was built on the function level similar to ours, but they consider only the complexity metrics.
% calculated by Understand C++.\footnote{\vspace{-5pt}\url{https://scitools.com/}}
They concluded that vulnerable functions have distinctive characteristics separating them from ``non-vulnerable but faulty'' functions.
They studied the JavaScript Engine from the Mozilla application framework. 
In this paper, we use 35 different metrics as predictors and build our prediction models specifically for JavaScript programs.

Shin~et~al.~\cite{shin2011evaluating} performed an empirical case study on two large code bases: Mozilla Firefox and Red Hat Enterprise Linux kernel, investigating if software metrics can be used in vulnerability prediction.
They considered complexity, code churn, and developer activity metrics.
The results showed that the metrics are discriminative and predictive of vulnerabilities.
%On Mozilla's code base, their model predicted 70.8\% of known vulnerabilities, and on Red Hat's Linux kernel it achieved 68.8\%.
However, their model was also built on file level, while we are predicting vulnerable functions.

Chowdhury~et~al.~\cite{chowdhury2011using} created a framework that can predict vulnerabilities mainly relying on the CCC (complexity, coupling, and cohesion) metrics~\cite{chidamber1994metrics}.
They also compared four statistical and machine learning techniques (namely C4.5 Decision Tree, Random Forests, Logistic Regression, and Naive Bayes classifier).
%Their case study was also built on the code-base of Mozilla Firefox and used file-level granularity.
%The created model was really accurate (with precision above 90\%), but the recall was only 20\% or lower which means an F-measure of 0.33 at most.
The authors concluded that decision-tree-based techniques outperformed statistical models in their case.
We also found that tree-based classifiers perform well for vulnerable JavaScript function prediction.

Morrison~et~al.~\cite{Morrison2015ChallengesWA} built a model -- replicating the vulnerability prediction model by Zimmermann~et~al~\cite{zimmermann2010searching} -- for both binaries and source code at file level.
%They figured out that vulnerability prediction at binary level is not actionable as it would take too much time to inspect the flagged binaries.
%Their results showed a precision below 0.5 and recall below 0.2.
The authors checked several learning algorithms including SVM, Naive Bayes, random forests, and logistic regression.
On their dataset, Naive Bayes and random forests performed the best.
In our setup, the Naive Bayes algorithm was the worst performer, while random forest achieved good results.

Yu~et~al. introduced HARMLESS~\cite{yu2018harmless}, a cost-aware active learner approach to predict vulnerabilities.
They used a support vector machine based prediction model with under-sampled training data, and a semi-supervised estimator to estimate the remaining vulnerabilities in a code base.
HARMLESS suggests which source code files are most likely to contain vulnerabilities.
They also used Mozilla's code base in their case study, with 3 different feature sets: metrics, text, and the combination of text mining and crash dump stack traces.
The same set of source code metrics were used than that of Shin~et~al.~\cite{shin2011evaluating}.
%They outperformed their previous study's \cite{theisen2015strengthening} recall and cost value.

All the above works target file-level vulnerability prediction, while we address the prediction of vulnerable JavaScript functions in our current work.

\vspace{-4pt}
\section{Approach}
\label{sec:methodology}
\vspace{-4pt}

\subsection{Dataset collection method}\label{sec:dataset_approach}

To build machine learning models, we needed a training dataset with features of JavaScript functions manually labeled as vulnerable or non-vulnerable.
The overview of the data mining process we performed is shown in Figure~\ref{fig:process_overview}.

\subsubsection{Processing nsp and Snyk and linking them with GitHub}\label{sec:dataset_process}

We leveraged two publicly available vulnerability databases, nsp (the Node Security Platform, which is now part of npm)~\cite{nspweb} and the Snyk Vulnerability Database~\cite{snyk_vuldb}.
Both of these projects aim to analyze programs for vulnerable third party module usages.
%While nsp targets Node.js programs specifically, Snyk supports various languages (e.g. Python, Java, JavaScript) and dependency management systems (e.g. pip, Maven, npm).
They have command line and/or web-based interfaces, which can inspect an arbitrary Node.js module to find external dependencies with known vulnerabilities.
To achieve this, they utilize a list of known vulnerabilities to look for security issues in the particular version of an external module the programs depend on.
We extracted and processed these vulnerability databases.

As for nsp, we used its command line interface to collect vulnerability data.
It provides a \emph{gather} command that saves its internal list of vulnerabilities into a JSON file.
Snyk has an online repository of known vulnerabilities, but there is no possibility for downloading the entries.
Nonetheless, Snyk maintains a GitHub mirror\footnote{\url{https://github.com/snyk/vulnerabilitydb}\vspace{-10pt}} of its vulnerability database with monthly synchronization.
We used the content of this GitHub repository in case of Snyk (accessed on 27/05/2018).

The main issue with these extracted raw vulnerability sources is that they contain unstructured data.
The entries include human readable description of vulnerabilities with URLs of fixing commits, pull requests or issues in GitHub or other repositories.
However, these URLs are somewhat arbitrary, they can appear on multiple places within the entries and any of them might be missing entirely.
To handle this, we wrote a set of Python scripts to process these vulnerability entries and create an internal augmented and structured representation of them.
The scripts collected all the URLs from each entry $vuln_i$ and kept all those pointing to GitHub.
Traversing these URLs, we derived a set of fixing commits (commits pointing to the state of the system where a security vulnerability has been fixed, thus they already contain the mitigation code) for each $vuln_i$ using the GitHub REST API following these steps:
\begin{enumerate}
	\item If the URL pointed to a particular commit, we put the appropriate commit hash to the fixing list ($fix_i$)
	\item If the URL pointed to a pull request or merge request, we put all the commit hashes in the request to $fix_i$
	\item If the URL pointed to an issue, we traversed through the comments of the issue and collected all mentioned URLs into a separate list for manual validation.
\end{enumerate}
\vspace{-2pt}
If the separate list for manual validation was not empty, we manually checked all the commit URLs in it and put only those commits into $fix_i$ that were indeed related to the fix of the original vulnerability issue ($vuln_i$).
The manual validation was performed by one of the authors, while another author participated in the discussion and resolution of problematic cases.
The added commits usually introduced unit tests or some corrections if the first fix was incomplete.

We note that it is possible that a commit which was referenced in the dataset entry (i.e. $fix_i$) contained tangled code changes (i.e. pull or merge requests).
To lower the risk, we performed a random cross-check on several of these large commits, but found no tangled changes in our sample.
%, so the likelihood of these tangled changes within fixes is low.

Upon finalizing the fixing commit lists for each entry, we collected all their code modifications in the form of a combined patch file ($patch_i$) that contained all the modifications from the fixing commits.
We used the GitHub API again to collect this information.
Moreover, we identified the parent commit of the first commit in time belonging to the vulnerability fix ($sha_{pre}$) for each system.
Version $sha_{pre}$ was used to assign the labels 1 or 0 to functions indicating if the function contained a vulnerability or not.
The final dataset was assembled from all the $sha_{pre}$ versions of the functions in the systems.
We marked all functions that were affected by any of the vulnerability fixing modifications (i.e. $patch_i$ changed those functions) as vulnerable.
All the other functions of the JavaScript programs were marked as non-vulnerable.
We note that all the test functions (i.e. functions contained in files under ``test'' folders) have been filtered out as these would only distort the prediction models.
%\todoi{ide kellene 1-2 gondolat, hogy tipikusan hany fuggvenyt erintett egy patch. Feltételezem, hogy 90\%-ban egyet(?). Esetleg johetne ide egy eloszlas hisztogram is: x tengelyen az erintett fuggvenyek szama, y oszlopmagassag a patch darabszam.}
%The technical details on how we mapped which functions were affected by a patch file and how we identified functions between %versions are described in Section~\ref{sec:js_name_matching}.

\subsubsection{Mapping patches with JavaScript functions}\label{sec:js_name_matching}

%In this section, we describe the details on how we mapped which functions were affected by a patch file.
To perform the mapping of patches to functions, we used the patch files ($patch_i$ for each vulnerability $vuln_i$ collected by the process described in Section~\ref{sec:dataset_process}) of the vulnerability-fixing commits in a unified diff format.
Each diff contains a header information specifying the name of the original and the new files.
After that, there are one or more change hunks that contain the actual line differences and each hunk begins with range information about the modification.
%By default, each hunk includes three unchanged lines before and after the actual change and these extra lines are included in the range showed at the beginning of the hunks.
%To avoid any false mapping of changes, we use diff files without these unchanged lines (by overriding the default setting).
We checked whether any function falls into this range.
We achieved this by using the source code positions of the functions -- begin and end line numbers, which were produced by the OpenStaticAnalyzer tool -- and checked whether these two ranges intersect or not.
An example is shown in Listing~\ref{lst:patch}.

\vspace{-13pt}
\begin{lstlisting}[language=diff, caption=Example diff file, label=lst:patch, firstnumber=1, xleftmargin=1.2cm]
--- /path/to/original.js    timestamp
+++ /path/to/new.js    timestamp
@@ -4,1 +4,2 @@
+  var tmp = bar(i);
+  return tmp;
-  return bar(i);
\end{lstlisting}
\vspace{-8pt}
\begin{lstlisting}[language=JavaScript, caption=Example JavaScript function, firstnumber=1, xleftmargin=1.2cm]
function foo(a) {
  var i = 4 * a;
  // call bar
  var tmp = bar(i);
  return tmp;
}
\end{lstlisting}
\vspace{-5pt}

The source position of the \textit{foo} function is [1,6] and the range from the diff is [4,5].
They intersect, so our method incorporates the \textit{foo} function into the dataset.
With this algorithm, we found all the functions that were changed by each vulnerability fixing commit, which we mapped to their previous versions (in $sha_{pre}$) to mark them vulnerable in the version prior to the first fixing commit.

\subsubsection{Static source code metrics}

For predictors (or, features), we used static source code metrics.
We calculated the metrics for the functions included in the final dataset using two tools, escomplex~\cite{escomplexweb} and OpenStaticAnalyzer (OSA)~\cite{osaweb}.
Both OpenStaticAnalyzer~\cite{DBLP:conf/icsoft/PengoG18} and escomplex~\cite{Chatzidimitriou:2018:NIM:3196398.3196465, phdthesis_Mariano} were used and referenced in related research works, thus we consider them to be reliable.
The list of calculated metrics is shown in Table~\ref{tab:metrics}.
Please note that similar metrics are grouped together in one line, so the total number of calculated metrics is 35.

%We mapped the OSA and escomplex metrics based on the line information of functions.
%Once we identified the vulnerable functions extracted from vulnerability fixing patches in version $sha_{pre}$ (see Section~\ref{sec:js_name_matching}), we used line information again to match them with the results of static analysis.

\subsubsection{Dataset structure}\label{sec:dataset_struct}

The final dataset structure follows a simple CSV format that is easy to feed into many machine learning frameworks.
Each line of the CSV file represents a function from a Node.js program.
The $1^{st}$ column is a short name, while the $2^{nd}$ is the qualified name of the function generated by the algorithm described in Section~\ref{sec:js_name_matching}.
The $3^{rd}$ column shows the path of the containing JavaScript source file, while the $4^{th}$ column contains a GitHub URL to the analyzed JavaScript source file (in the $sha_{pre}$ version).
% Table generated by Excel2LaTeX from sheet 'Munka1'
\begin{table}[htbp]
  \centering
	\setlength\tabcolsep{5.5pt}
	\scriptsize
  \caption{Calculated static source code metrics}
	\vspace{-5pt}
    \begin{tabular}{l|ll}
    \textbf{Metric} & \textbf{Description} & \textbf{Tool} \\
    \midrule
    \midrule
    CC    & Clone Coverage & OSA \\
    CCL   & Clone Classes & OSA \\
    CCO   & Clone Complexity & OSA \\
    CI    & Clone Instances & OSA \\
    CLC   & Clone Line Coverage & OSA \\
    LDC   & Lines of Duplicated Code & OSA \\
    McCC, CYCL & Cyclomatic Complexity & OSA, escomplex \\
    NL    & Nesting Level & OSA \\
    NLE   & Nesting Level without else-if & OSA \\
    CD, TCD & (Total\footnote{Total means that the metric is calculated for the actual code element including all the contained elements recursively.\vspace{-12pt}}) Comment Density & OSA \\
    CLOC, TCLOC & (Total) Comment Lines of Code & OSA \\
    DLOC  & Documentation Lines of Code & OSA \\
    LLOC, TLLOC & (Total) Logical Lines of Code & OSA \\
    LOC, TLOC & (Total) Lines of Code & OSA \\
    NOS, TNOS & (Total) Number of Statements & OSA \\
    NUMPAR, PARAMS & Number of Parameters & OSA, escomplex \\
    HOR\_D & Nr. of Distinct Halstead Operators & escomplex \\
    HOR\_T & Nr. of Total Halstead Operators & escomplex \\
    HON\_D & Nr. of Distinct Halstead Operands & escomplex \\
    HON\_T & Nr. of Total Halstead Operands & escomplex \\
    HLEN  & Halstead Length & escomplex \\
    HVOC  & Halstead Vocabulary Size & escomplex \\
    HDIFF & Halstead Difficulty & escomplex \\
    HVOL  & Halstead Volume & escomplex \\
    HEFF  & Halstead Effort & escomplex \\
    HBUGS & Halstead Bugs & escomplex \\
    HTIME & Halstead Time & escomplex \\
    CYCL\_DENS & Cyclomatic Density & escomplex \\
    \end{tabular}%
		\vspace{-17pt}
  \label{tab:metrics}%
\end{table}%

The $5^{th}$ and $6^{th}$ columns contain the starting, while the $7^{th}$ and $8^{th}$ the ending line and column information, respectively.
Columns 9 to 43 contain the calculated metric values listed in Table~\ref{tab:metrics}.
The last column (column 44) contains the flag indicating whether the function is vulnerable or not.

The created vulnerability dataset\footnote{\url{http://www.inf.u-szeged.hu/~ferenc/papers/JSVulnerabilityDataSet/}} consists of 12,125 JavaScript functions from which 1,496 are vulnerable.
%It is publicly available for download.

%\input{data/alg_params}

\subsection{Dataset analysis approach}\label{sec:approach}

We employed 8 different types of machine learning algorithms on the vulnerability dataset created with the method described in Section~\ref{sec:dataset_approach}.
These algorithms were two deep neural network variants, a simple (DNN$_s$) and a complex one (DNN$_c$), the K-Nearest Neighbors algorithm (KNN), a decision tree classifier (Tree), the C-Support Vector Classification variant of Support Vector Machine algorithm (SVM), Random Forest (Forest), Logistic regression (Logistic), Linear regression (Linear) and the Gaussian Naive Bayes algorithm (Bayes).
The deep neural network algorithms were implemented in the \emph{TensorFlow}~\cite{abadi2016tensorflow} framework\footnote{\url{https://www.tensorflow.org/}}, while we used \emph{scikit-learn}\footnote{\url{http://scikit-learn.org/stable/}\vspace{-13pt}} to run all the other algorithms.
Both frameworks were used in a Python environment.
We could not use only one of them because while TensorFlow has a strong support for deep learning, it does not contain all the classic algorithms.
In contrast, scikit-learn is very strong in classic machine learning algorithms but it is not a deep learning framework in itself.

DNN$_s$ stands for the base DNN algorithm implemented in TensorFlow.
We used it without any modifications except for changing the parameters it provides (see Section~\ref{sec:param_search}).
DNN$_s$ learning runs for a fixed number of iterations over all the training instances (i.e. epochs).
DNN$_c$ is our own modified strategy for training a DNN.
It uses an adaptive learning rate method where the learning rate parameter is not constant over the course of training.
We start with a relatively high learning rate parameter and continue the classic back propagation algorithm until there is no improvement in the value of F-measure (we call this a \emph{miss}).
Then we reduce the learning rate parameter to half, restore the previous model state, and continue the learning process from there.
We repeat these steps until we get 4 misses in succession, then terminate the algorithm and return the last, best performing model.
This strategy reduces the likelihood of the algorithm getting ``stuck'' in a local optimum.
Regarding KNN, Tree, SVM, Forest, Logistic and Linear regression, and the Naive Bayes algorithm, we used their scikit-learn implementation.

\subsubsection{Grid search for the best parameters}\label{sec:param_search}

To find the best performing configuration of each algorithm, we applied a grid search approach~\cite{bergstra2011algorithms} on the hyper parameters of the learning algorithms.
It means that we defined various values for machine learning algorithm parameters and trained multiple models using various combinations of hyper parameters.
After having multiple results for each model, we could select the best performing ones.

\looseness=-1
For all training sessions we divided the training data into three sets, \emph{train}, \emph{dev}, and \emph{test} in a 80\%, 10\%, 10\% proportion, respectively, and used a 10-fold cross-validation.
%We iterated the test and dev parts over all the training instances and collected an aggregated confusion matrix.
%This confusion matrix contained the sum of true positive, true negative, false positive and false negative hits through all 10 folds.
At the end of the 10 folds, we calculated the precision, recall, and F-measure values. % from these aggregated confusion matrices.
For selecting best performing parameter configurations, we relied only on the results of the dev set.
This ensured that we did not use information for selecting the best parameters from our final test set in any way.
We used F-measure as our primary performance indicator as in the security domain both precision and recall are important.
%The particular parameters and their possible values we checked for each algorithm is summarized in Table~\ref{tab:params}.
%We trained each algorithm using the combination of parameter values indicated within the square brackets.

%The value ``-'' means the parameter is optional, thus we left them out from some combinations.
%Linear regression and Naive Bayes algorithm does not support any hyper parameters, thus they were run in only one configuration.
%For the full reference of these parameters, please check the TensorFlow\footnote{\url{https://www.tensorflow.org/api_docs/python/tf/contrib/learn/DNNClassifier}} and scikit-learn\footnote{\url{http://scikit-learn.org/stable/user_guide.html}} technical documentations.

\subsubsection{Sampling strategies}\label{sec:sampling-strategy}

%Another aspect we analyzed in connection with vulnerability prediction is the impact of re-sampling strategies on the training.
%It is very common in bug and vulnerability prediction that the initial training dataset is highly imbalanced.
%It means that there are significantly more training instances for one class (typically bug-free or non-vulnerable code parts) than for the other (typically buggy or vulnerable components).
In our assembled vulnerability dataset, only slightly more than 10\% of the functions were marked as vulnerable.
This highly imbalanced nature of the training set is usually unwanted as prediction models might be distorted by these skewed distributions.

\looseness=-1
A common way of handling such situations is the usage of random under or over-sampling strategies~\cite{Batista:2004:SBS:1007730.1007735}.
Random under-sampling means we randomly throw away training instances from the larger set until we reach a pre-defined ratio between the two classes.
Random over-sampling is when we randomly repeat training instances from the smaller set until we reach a pre-defined ratio between the two classes.
%Both strategies achieve a more balanced training set artificially, but they affect the final prediction model performances differently.

We repeatedly ran our algorithm parameter grid search (see Section~\ref{sec:param_search}) with the following re-sampling strategies: \emph{no re-sampling} (None); \emph{over-sampling} ($\uparrow$) with ratios 25\%, 50\%, 75\% and 100\%; \emph{under-sampling} ($\downarrow$) with ratios 25\%, 50\%, 75\% and 100\%.
%For instance, a 50\% over-sampling strategy means that we randomly repeat training instances from the vulnerable class until we reach 50\% in number compared to the training instances in the non-vulnerable class.
%Analogously, a 50\% under-sampling strategy means we leave out training instances from the non-vulnerable class until it contains only twice as many instances as the vulnerable class.
%It is important to note here that while over-sampling does not result in loss of training data, under-sampling reduces the number of used training samples.

\vspace{-3pt}
\section{Results}
\label{sec:results}
\vspace{-3pt}

We trained 9 different prediction models on the created vulnerability dataset (8 different algorithms, but two variants of DNN)
%We used a classification with two classes on JavaScript functions: vulnerable or not.
on a desktop PC\footnote{8 core 2.4GHz CPU, NVIDIA Titan Xp GPU, 8GB RAM\vspace{-10pt}} using both CPU and GPU.
The running times varied between 6-12 hours for a complete hyper-parameter grid-search of all algorithms.
We repeated these grid-search sessions for all the separate over and under-sampling strategies (described in Section~\ref{sec:sampling-strategy}), thus building all the models took a considerable amount of time and computing resources.

\vspace{-4pt}
\subsection{Results on the imbalanced dataset}
\begin{figure}
\centering
\includegraphics[width=0.9\columnwidth]{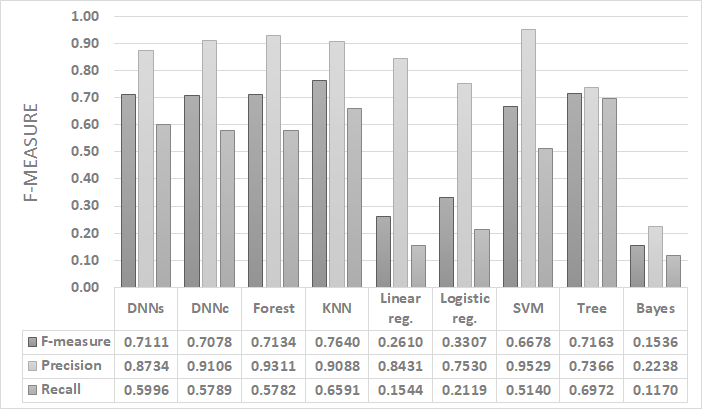}
\vspace{-5pt}
\caption{Results on the imbalanced dataset}
\vspace{-20pt}
\label{fig:results_overview}
\end{figure}

\vspace{-1pt}
First, we ran our grid-search without applying any re-sampling on the vulnerability dataset, which is highly imbalanced (out of 12,125 functions only 1,496 are vulnerable).
The performances of the 9 models with their best parameter combinations is displayed in Figure~\ref{fig:results_overview}.

The overall results are surprisingly good given the fact that JavaScript is a highly dynamic language and we used only static source code metrics as predictors.
Five out of the 9 models (DNN$_s$, DNN$_c$, Forest, KNN, and Tree) achieved an F-measure of over 0.70 and SVM was also very close with 0.67.
It is interesting to note that for all algorithms, precision values were significantly higher than recall, except for the decision tree classifier, which had a precision of 0.74, a recall of 0.7 and an F-measure of 0.72.

\looseness=-1
Only the Naive Bayes algorithm was clearly incapable of producing a viable prediction model using the original, imbalanced vulnerability dataset.
Logistic and linear regression achieved a precision of 0.75 and 0.84, respectively, which are relatively high values, however, they had a very low recall (0.21 and 0.15, respectively) that decreased the F-measure values.

As a simple baseline, we also ran the ZeroR algorithm with the setup that it predicted all instances to be vulnerable (and not vice versa as the default setup would do, because if we predicted every instance to be non-vulnerable, all our IR metrics would have been 0).
ZeroR achieved a precision of 0.12 and perfect recall of 1 (as it found all the vulnerable instances), which adds up to an F-measure of 0.21.
This result is much worse than those of the other algorithms' except for the Naive Bayes.
Therefore, we can already answer RQ1 based on these results.
\smallskip

\noindent
\fbox{
\parbox{0.94\columnwidth}{
\textbf{RQ1:} Choosing a suitable algorithm with proper parameters, it is possible to create efficient function level vulnerability prediction models using only static source code metrics as predictors. The DNN, KNN, Forest, and Tree algorithms all achieved F-measures above 0.70 without any re-sampling on the dataset.
}
}

% Table generated by Excel2LaTeX from sheet 'aggr_f'
\begin{table}[htbp]
  \centering
	\scriptsize
	\setlength\tabcolsep{1pt}
		\caption{F-measures achieved by the machine learning algorithms}
	\vspace{-6pt}
    \begin{tabular}{l|rrrrrrrrrr}
    \textbf{Alg.} & \multicolumn{1}{l}{\textbf{None}} & \multicolumn{1}{l}{\textbf{$\uparrow$25\%}} & \multicolumn{1}{l}{\textbf{$\uparrow$50\%}} & \multicolumn{1}{l}{\textbf{$\uparrow$75\%}} & \multicolumn{1}{l}{\textbf{$\uparrow$100\%}} & \multicolumn{1}{l}{\textbf{$\downarrow$25\%}} & \multicolumn{1}{l}{\textbf{$\downarrow$50\%}} & \multicolumn{1}{l}{\textbf{$\downarrow$75\%}} & \multicolumn{1}{l}{\textbf{$\downarrow$100\%}} & \multicolumn{1}{l}{\textbf{Rand}} \\
    \midrule
    \midrule
    DNN$_s$ & 0.71 & \textbf{0.71$^*$} & 0.71 & 0.65 & 0.68 & 0.70 & 0.71 & 0.69 & 0.59 & 0.05 \\
    DNN$_c$ & 0.71 & 0.70 & 0.71 & 0.68 & 0.65 & \textbf{0.71$^*$} & 0.71 & 0.68 & \cellcolor[rgb]{ .651,  .651,  .651} \textbf{0.66} & 0.01 \\
    Forest & 0.71 & \textbf{0.74$^*$} & \cellcolor[rgb]{ .651,  .651,  .651} \textbf{0.74} & \cellcolor[rgb]{ .651,  .651,  .651} \textbf{0.73} & \cellcolor[rgb]{ .651,  .651,  .651} \textbf{0.72} & 0.72 & 0.72 & 0.72 & 0.65 & 0.05 \\
    KNN   & \cellcolor[rgb]{ .651,  .651,  .651} \textbf{0.76$^*$} & \cellcolor[rgb]{ .651,  .651,  .651} \textbf{0.75} & 0.72 & 0.6935 & 0.6817 & \cellcolor[rgb]{ .651,  .651,  .651} \textbf{0.76} & \cellcolor[rgb]{ .651,  .651,  .651} \textbf{0.75} & \cellcolor[rgb]{ .651,  .651,  .651} \textbf{0.74} & 0.64 & 0.14 \\
    Lin. reg. & 0.26 & 0.48 & \textbf{0.55$^*$} & 0.49 & 0.45 & 0.30 & 0.37 & 0.51 & 0.44 & 0.02 \\
    Log. reg. & 0.33 & 0.50 & \textbf{0.57$^*$} & 0.55 & 0.49 & 0.38 & 0.45 & 0.53 & 0.49 & 0.01 \\
    SVM   & 0.67 & 0.70 & \textbf{0.72$^*$} & 0.70 & 0.68 & 0.67 & 0.67 & 0.67 & 0.65 & 0.16 \\
    Tree  & \textbf{0.72$^*$} & 0.71 & 0.71 & 0.71 & 0.70 & 0.70 & 0.69 & 0.67 & 0.59 & 0.15 \\
    Bayes & 0.15 & 0.16 & 0.16 & \textbf{0.21$^*$} & 0.20 & 0.16 & 0.16 & 0.18 & 0.17 & 0.07 \\
    \midrule
    \textbf{Median} & 0.71 & 0.70 & \textbf{0.71$^*$} & 0.68 & 0.68 & 0.70 & 0.69 & 0.67 & 0.59 & 0.05 \\
    \bottomrule
    \end{tabular}%
  \label{tab:f-measure-overview}%
	\vspace{-20pt}
\end{table}%

\subsection{Comparison of the models based on the complete results}
\vspace{-3pt}
The best performing model results based on the complete grid-search using various re-sampling strategies are summarized in Table~\ref{tab:f-measure-overview}.
Each column of the table contains model results (in terms of F-measure\footnote{Matthews correlation coefficients (MCC) were slightly smaller in general, but they showed the same tendency, see the shared dataset for details.\vspace{-15pt}}) using a particular re-sampling strategy (see Section~\ref{sec:sampling-strategy}) with the best parameters found by the grid-search method.
The first column shows the results on the original imbalanced dataset without re-sampling (in line with Figure~\ref{fig:results_overview}).
The next four columns display the results on the over-sampled, while the following four on the under-sampled dataset.
The last column presents results on a random sanity check.
To make sure that having these strong prediction results is not coincidental, we created a new training dataset by reassigning the 1,496 vulnerable labels randomly.
The training results on this randomly labeled dataset shows that models cannot learn to distinguish arbitrary set of functions based on their static source code metrics, thus our prediction results are unlikely to be the consequences of random factors.

\begin{figure}[htb!]
\centering
\includegraphics[width=0.98\columnwidth]{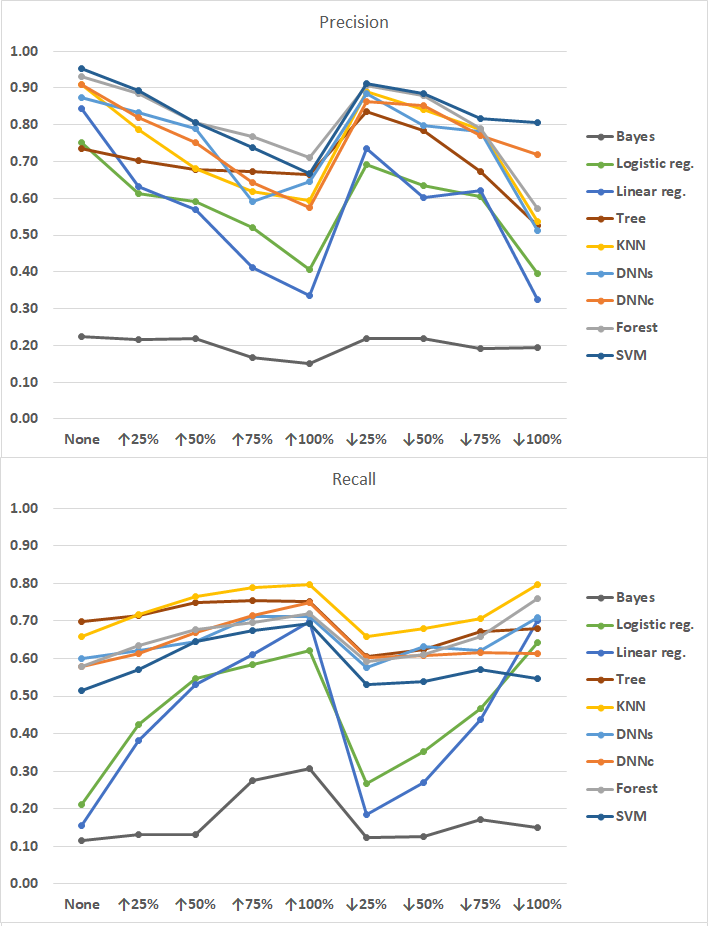}
\vspace{-5pt}
\caption{Impact of re-sampling on the learning precision and recall}
\label{fig:resample_prf}
\vspace{-20pt}
\end{figure}

\looseness=-1
The gray cells in the table mark the best performing algorithm with the given re-sampling strategy.
KNN is the best in five different re-sampling configurations, Forest in three, while DNN$_c$ in one.
The values indicated in bold and with an asterisk are the best F-measure values for a given machine learning algorithm (i.e. the highest value in the row).
The most important thing to note here compared to the results on the imbalanced training set is that even SVM achieved a result above 0.70 with an appropriate over-sampling strategy ($\uparrow$50\%, $\uparrow$75\%).
Seven out of the nine models achieved better performances in some of the re-sampling configurations than on the original, imbalanced dataset.
The exact composition of precision and recall values leading to this F-measures are visualized in Figure~\ref{fig:resample_prf}.
Based on the data in Table~\ref{tab:f-measure-overview} and Figure~\ref{fig:resample_prf}, we can answer RQ2 as follows.

%\subsection{The effect of re-sampling on model performances}\label{sec:sampling_effect}
%
%The variation of prediction performances caused by the different re-sampling strategies can be seen in Figure~\ref{fig:resample_prf}.

\vspace{2pt}
\noindent
\fbox{
\parbox{0.94\columnwidth}{
\textbf{RQ2:} The best performing algorithm for predicting vulnerable JavaScript functions in terms of F-measure was KNN with an F-measure of 0.76 (0.91 precision and 0.66 recall).
The best precision (0.95) was achieved by SVM, while the best recall (0.80) by KNN. %, but with a different re-sampling strategy than the one yielding the best F-measure.
In overall, KNN, DNN, SVM, Tree, and Forest are equally well-suited for the task, while the regressions as well as the Naive Bayes algorithm perform much worse.
}
}

\vspace{-8pt}
\section{Threats to Validity}\label{sec:threats}
\vspace{-2pt}

%In their work Basili~et~al.~\cite{basili1999building} stated that generalizing results from empirical studies in software engineering is difficult.
%This is because any process, any study depends on quite a lot of context variables.
%Therefore, our results might be affected by this phenomenon as well and might change with context (e.g. building vulnerability prediction models for other languages).
%However, in the current context (i.e. JavaScript vulnerable function prediction) the results rely on an extensive dataset and robust machine learning approaches with lots of various training setups.

Our data collection process might not be 100\% accurate as only the additional candidate commits collected from issue comments were validated manually.
The original data sources might contain errors as well as our automatic patch collection and patch-to-function mapping algorithms might introduce inconsistencies.
We tried to mitigate this problem by thorough code review of our scripts and programs.

We mapped static source code analysis results of various tools and functions identified in patches by line information.
This is another source of possible errors, but we performed a small evaluation on 20 randomly selected JavaScript functions from the dataset and found no multiple functions in the same line.
Based on this and our past experience, we believe it is a safe assumption that multiple functions in the same code line are very rare in a non-minified JavaScript program.
As we used line information only within the same version of the programs, the likelihood of mismatching functions is even more negligible.

The extraction of features (i.e. static source code metrics) is heavily dependent on the accuracy of the tools used, which may threaten the extraction process.
However, there are numerous related works using the same tools, thus they can be considered stable.
Moreover, we manually double-checked some of the calculated metric values and found no problems in their calculation.

%\vspace{-5pt}
\section{Conclusions and Future Work}\label{sec:conclusions}
\vspace{-2pt}

In this paper, we published a novel JavaScript vulnerability dataset to be used for building prediction models.
The dataset contains various JavaScript functions together with their static source code metrics and a flag indicating whether the function contains a vulnerability or not.
This information was assembled by mining public vulnerability data sources of nsp and Snyk and collecting fixing patches from GitHub.

We presented an assessment of existing machine learning algorithms for building function level vulnerability prediction models using this dataset.
We analyzed the performances of 8 different types of algorithms using the training set as is, and also by applying various re-sampling strategies.
%We found the best hyper-parameters of the algorithms using grid-search.

Our results show that even for such a highly dynamic language as JavaScript, static source code metrics are suitable predictors of vulnerabilities.
However, we experienced large variances in prediction performances depending on the applied sampling strategy and hyper-parameters.
Using the appropriate machine learning algorithm (DNN, KNN, Tree, Forest, or SVM) and suitable hyper-parameters, a prediction with F-measure of 0.7 and above can be achieved.
Nonetheless, there is a clear trade-off between precision and recall; over-sampling tends to improve recall, but decreases precision, while intensive under-sampling improves precision, but reduces recall significantly.
%Thus, selecting the appropriate algorithm, sampling method, and hyper-parameters is a complex task, and one needs to consider the desired trade-off in terms of precision and recall.

%az alábbi két mondatot kivettem, nem igazán illik ide, mivel bug-okról szól, és hivatkozás is kellene, ami alátámasztja
%We are well aware that bug prediction results clearly show the superiority of version control history metrics for prediction over static source code metrics.
%Additionally, the nature of vulnerabilities differ from bugs, thus text-like approaches might as well be viable alternatives.
We plan to extend the set of predictors with history and textual metrics in order to further improve vulnerability prediction at the level of JavaScript functions.

\vspace{-4pt}
\section*{Acknowledgment}
\vspace{-2pt}
The research has been supported by the National Research, Development and Innovation Fund of Hungary, financed under the 2018-1.2.1-NKP funding scheme.
Ministry of Human Capacities, Hungary grant 20391-3/2018/FEKUSTRAT is acknowledged.
The Titan Xp used for this research was donated by the NVIDIA Corporation.
\vspace{-17pt}
\bibliographystyle{./IEEEtran}
\bibliography{bibl}

\end{document}